\documentstyle[12pt,epsfig]{article} 
\begin{document}
\setlength{\parskip}{0.45cm}
\setlength{\baselineskip}{0.75cm}
\begin{titlepage}
%
\vspace{0.6cm}
\begin{center}
\Large
 { \bf  Phenomenological Analysis of Data on Inclusive and 
Semi-inclusive Spin Asymmetries}
\vspace{1.2cm}

\large 
 M. Kurzela, J. Bartelski
 \vspace*{0.4cm}

\normalsize
{\it 
Institute of Theoretical Physics, Warsaw University,
  Hoza 69, 00-681 Warsaw, Poland}
 \vspace*{1.5cm}

\large 
  S. Tatur
 \vspace*{0.4cm}

\normalsize
{\it 
Nicolaus Copernicus Astronomical Center, Polish Academy of Sciences,
Bartycka 18, 00-716 Warsaw, Poland
}
\vspace*{1.5cm}

\large
{\bf Abstract}
\end{center}
\vspace{0.2cm}
\normalsize

 We present a phenomenological  analysis of data on both inclusive and
semi-inclusive spin asymmetries. We examine the impact of the  
semi-inclusive results presented by SMC on the determination of
polarized parton distributions performing global fits
with different sets of observables. We discuss the flavour dependence
of the polarized sea inside a nucleon.

\begin{flushleft}
July 1998
\end{flushleft}
\end{titlepage}
\newpage

In recent years  a number of theoretical attempts 
\cite{GRSV,GS,ABFR,E154th,BTincl,BT,deF,Str,LSS,Bourr} to determine the 
polarized quark parton distributions in the nucleon have been performed.
The deep inelastic polarized structure functions $g^N_1(x, Q^2)$ 
or the asymmetries  measured in inclusive processes 
\cite{E80,E130,EMC,E142,E143p,E143d,E143Q2,SMCp,SMCd,E154,HERMES} are used in
phenomenological analyses. Such an analysis of the first moment of 
the structure function $\Gamma ^N_1=\int_0^1 g^N_1(x) dx$ pointed out that
quarks carry little of the spin of the nucleon \cite{EMC,BEK,AR}. 
The reasonable  suggestion is that the sea quarks and/or gluons are polarized. 
However, inclusive deep inelastic scattering does not provide sufficient 
information about the flavour separation of the polarized sea. Hence different
combination of the polarized parton distributions have to be measured 
in order to get more 
information about flavour structure of the polarized sea.

The measurement of the semi--inclusive spin asymmetries for positively and 
negatively charged hadrons from deep inelastic scattering of polarized muons 
on polarized protons and deuterons provides additional data on required 
observables. Presently available semi-inclusive results 
\cite{EMCsi,SMCsi96,SMCsi97} can be used to 
determine polarized valence and non--strange sea quark distributions, 
independently from totally inclusive data. The aim of the paper is to 
combine two 
kinds of existing data, inclusive and semi--inclusive, to extract polarized 
parton distributions.

  Measurement of the inclusive deep inelastic lepton nucleon scattering gives 
information about the spin asymmetry \cite{Ans}:
\begin{equation} \label{eq0}
A_{1}^{N}(x, Q^2) \simeq
\frac{g_{1}^{N}(x, Q^2)}{F_{1}^N(x, Q^2)} ,
\end{equation}
 which in leading order QCD parton model is given by:
\begin{equation} 
A_{1}^{N}(x, Q^2) \simeq
\frac{\sum_q e_q^2 \Delta q(x, Q^2)}{\sum_q e_q^2 q(x, Q^2)},
\end{equation}
where $e_q$ is the charge of the $q$--flavoured quark, $q$ and $\Delta q$
denotes unpolarized and polarized quark distributions respectively,
where $q=u, d, s,\bar{u},\bar{d},\bar{s}$. This is the consequence of the fact 
that in LO QCD $2 g_1=\sum_q e_q^2 \Delta q$, 
$\frac{1}{x}F_2=\sum_q e_q^2 q$ and of the Callan--Gross relation $F_2=2xF_1$.
First attempt to improve such a model is to use 
$R^N(x,Q^2)=(F^N_2-2 x F^N_1)/2 x F^N_1 \not=0$, which is the ratio of the absorption
cross--sections for virtual longitudinal and transverse photons 
($R=\sigma_L/\sigma_T$) \cite{Rth}. In calculations, we use the
parametrization of $R$ described in Ref.\cite{BKK}, 
which is analogous to the one
given in  Ref.\cite{W90}, but fitted to the enlarged set of 
data on $R$ with new experimental values \cite{Dasu,NMC,E140X,CDHSW}. This 
correction leads to the expression:
\begin{equation} \label{eq2} 
A_{1}^{N}(x, Q^2) \simeq
\frac{\sum_q e_q^2 \Delta q(x, Q^2)}{\sum_q e_q^2 q(x, Q^2)}
(1+R^N(x,Q^2)) ,
\end{equation}
for proton and neutron target ($N=p,n$). The parton distributions are 
those of the proton whereas for neutron are obtained by the isospin 
interchange $u\leftrightarrow d$. For deuteron target case one has 
to multiply above 
expression by  additional factor $1-3/2 p_D$ where $p_D$ is a probability 
of D-state in deuteron wave function ($p_D=0.05\pm 0.01$) \cite{omegaD}.

Analogously, for the  semi-inclusive asymmetries, the expression in the same
order can be written as: 
  \begin{equation} \label{eq4}
\left.
A_{1}^{N\,h}(x, Q^2) \right|_Z \simeq \frac{\int_{Z} dz\,
g_{1}^{N\,h}(x,z,Q^2)}{\int_{Z} dz\,F_{1}^{N\,h}(x,z,Q^2)},
\end{equation} 
where $h$ denotes the hadron detected in the final state and the variable z
is given by $E_h/E_N(1-x)$ with energies given in $\gamma^*p$ CM frame.
The region $Z$ is determined by kinematical cuts in measurement of the 
asymmetries.
Summing over positively charged hadrons, i.e. $\pi^+$, $K^+$ and $p$, and 
negatively charged ($\pi^-$, $K^-$, $\bar{p}$) respectively, we get:
\begin{equation} \label{eq1}
A_{1}^{N+(-)}(x, Q^2)\simeq
\frac{\sum_{q,h^{+(-)}} e_q^2 \Delta q(x, Q^2) D^h_q(Q^2)}
     {\sum_{q,h^{+(-)}} e_q^2 q(x, Q^2) D^h_q(Q^2)}
(1+R^N(x,Q^2)) .
\end{equation}
Here $D^h_q(Q^2)=\int^1_{0.2} dz D^h_q(z,Q^2)$ and $D^h_q(z,Q^2)$ is the
fragmentation function which represents the probability that a struck quark
with a flavour $q$ fragments into a hadron $h$. To reduce the number of
independent fragmentation functions one can use charge invariance and
isospin rotation symmetry as well as assumption for the unfavoured and favoured
fragmentation \cite{EMCsi,SMCsi96}. Further assumption concerning the
strange quark fragmentation function (e.g. $D^{K+}_s + D^{K-}_s =2
D^{K+}_u$) reduces the number of independent fragmentation functions to 6.
Finally the set of different weights in eq.~(\ref{eq1}) is:
\begin{eqnarray} \nonumber
\sum_{h^+} D^h_u=\sum_{h^-} D^h_{\bar{u}}=
D^{\pi^+}_u + D^{K^+}_u +D^{p}_u \; , \;\;\;\;&&
\sum_{h^-} D^h_u=\sum_{h^+} D^h_{\bar{u}}=
D^{\pi^-}_u + D^{K^-}_u +D^{\bar{p}}_u\; ,\\
\nonumber
\sum_{h^+} D^h_d=\sum_{h^-} D^h_{\bar{d}}=
D^{\pi^-}_u + D^{K^-}_u +D^{p}_u \; , \;\;\;\;&&
\sum_{h^-} D^h_d=\sum_{h^+} D^h_{\bar{d}}=
D^{\pi^+}_u + D^{K^-}_u +D^{\bar{p}}_u\; ,\\
\sum_{h^+} D^h_s+\sum_{h^+} D^h_{\bar{s}}=2(D^{\pi^+}_u + 
D^{K^+}_u + D^{p}_u)\; . &&
\end{eqnarray}
The presence of different $\sum_h D^h_q$ in eq.~(\ref{eq1}) enables to
examine combination of the polarized parton distributions different than
in the inclusive case.

To compare theoretical predictions of eq.~(\ref{eq2}) and eq.~(\ref{eq1}) with
experimental results we have to construct or choose the set of unpolarized
and polarized quark parton distribution functions. These functions are
combinations of the elementary ones, i.e. density of quarks with spin
parallel to the nucleon spin $q^+(x,Q^2)$ and density of quarks with
spin anti-parallel to the nucleon spin $q^-(x,Q^2)$. In details: 
$q(x,Q^2)=q^+(x,Q^2)+q^-(x,Q^2)$ and 
$\Delta q(x,Q^2)=q^+(x,Q^2)-q^-(x,Q^2)$. Our assumption is that
distributions $q^+$ and $q^-$ have the same functional behaviour, so there is
the only difference in the numerical coefficients \cite{BT}. It is not
necessarily true for $q$ and $\Delta q$ because the appropriate
coefficients in $q^+$ and $q^-$ could be equal (or have the same absolute
value but opposite sign) and in this case equivalent coefficients in $q$
($\Delta q$) vanish. The idea is to use formulas for the unpolarized quark
parton distributions as an input, then to extract from them formulas for $q^+$
and $q^-$ distributions just by splitting the numerical constants.
 
Previously this idea was explored in Ref.\cite{BT}\footnote{It is not the
first paper on this subject (see references therein) but the data on 
semi--inclusive asymmetries was included into the fit for the first time.},
where the latest version of the MRS \cite{MRS} parametrization was used.
To test the dependence of final results on the input parametrization we
have chosen the latest version of GRV parametrization for unpolarized
parton distributions \cite{GRV94}.This parametrization gives for the 
valence quarks at $Q^2=4 GeV^2$:
\begin{eqnarray} \nonumber
u_v(x)&=&3.221x^{-0.436}(1-x)^{3.726}(1-0.689x^{0.2}+2.254x+1.261x^{\frac{3}{2}}),\\
d_v(x)&=&0.507x^{-0.624}(1-x)^{4.476}(1+1.615x^{0.553}+3.651x+1.3x^{\frac{3}{2}}),
\end{eqnarray}
whereas for the sea anti-quarks:
\begin{eqnarray} \nonumber
\bar{s}(x)&=&0.0034x^{-1}(1-x)^{6.166}(1-2.392\sqrt{x}+7.094x)
e^{2.592\sqrt{\ln \frac{1}{x}}} (\ln\frac{1}{x})^{-1.15}, \\
S(x)&=&x^{-1}(1-x)^{6.356}\left[0.00285
e^{2.003\sqrt{\ln\frac{1}{x}}}+\right.\\
\nonumber
&&\left.x^{0.158}(0.738-0.981x+1.063x^2)(\ln\frac{1}{x})^{0.037}\right],\\ 
\nonumber
\delta(x)&=&0.107x^{-0.596}(1-x)^{8.621}(1+0.441x^{0.876}+18.721x),
\end{eqnarray}
where $S(x)=\bar{d}(x)+\bar{u}(x)$ is the non-strange singlet contribution to
the sea and $\delta(x)=\bar{d}(x)-\bar{u}(x)$ is the isovector non-strange
part of the quark sea. For the unpolarized gluon distribution we get:
\begin{equation}
G(x)=x^{-1}(1-x)^{5.566}\left[ 0.0527 e^{2.141\sqrt{\ln \frac{1}{x}}}+
x^{0.731}(5.11-1.204x-1.911x^2)(\ln\frac{1}{x})^{-0.472}\right].
\end{equation}
Generally the unpolarized parton distribution for the valence quarks and
the isovector non-strange part of the sea can be written in the form
$q(x)=x^{\alpha_q}(1-x)^{\beta_q}W(x)$. In other cases there is a similar
part $(1-x)^{\beta_q}$, which describes asymptotic behaviour for $x$ tending
to~1 but terms responsible for behaviour for $x$ tending to~0 are
more complicated.

Now we split the above distribution functions between $q^+$ and $q^-$ 
in order to get polarized parton distributions as $\Delta 
q(x)=q^+(x)-q^-(x)$. The asymptotic behaviour for $x\rightarrow 1$ (i.e. 
the value of $\beta_q$) is the same for all distributions like in the
unpolarized case and for $x\rightarrow 0$ (the value of $\alpha_q$) it
remains unchanged for valence quarks and the isovector part of quark sea. We
must be more careful in treating the strange sea and the isoscalar part.
Assuming that the polarized structure function $g_1$ have to be integrable
one has to split appropriate numerical constants in such
a manner that non-integrable terms of unpolarized parton distributions
disappear in polarized parton distributions (i.e. one has to split these
coefficients equally between $q^+$ and $q^-$). This procedure, of course,
changes asymptotic behaviour but functions remain integrable despite singular
behaviour at $x\rightarrow 0$. Our expressions for $\Delta q(x)$ are:
\begin{eqnarray} \label{eq10}  \nonumber
\Delta u_v(x)&=&x^{-0.436}(1-x)^{3.726}(A_u+B_u~x^{0.2}+C_u~x+D_u~x^{\frac{3}{2}}),\\
\nonumber
\Delta d_v(x)&=&x^{-0.624}(1-x)^{4.476}(A_d+B_d~x^{0.553}+C_d~x+D_d~x^{\frac{3}{2}}),\\
\nonumber
\Delta\bar{s}(x)&=&x^{-0.5}(1-x)^{6.166}(A_s+B_s \sqrt{x})
e^{2.592\sqrt{\ln \frac{1}{x}}} (\ln\frac{1}{x})^{-1.15}, \\
\Delta S(x)&=&x^{-0.842}(1-x)^{6.356}
(A_S+B_S~x+C_S~x^2)(\ln\frac{1}{x})^{0.037},\\ 
\nonumber
\Delta\delta(x)&=&x^{-0.596}(1-x)^{8.621}(A_{\delta}+B_{\delta}x^{0.876}),
\end{eqnarray}
where we have introduced 15 new parameters.

The polarized parton distribution functions must satisfy positivity
constraint,
\begin{equation}
\left| \Delta q(x,Q^2)\right| \leq q(x,Q^2),
\end{equation}
which leads to several constraints on coefficients in each distribution.
Furthermore we fix the normalization of the non-singlet distributions 
using the experimental value of the axial charge:
\begin{equation} \label{eq3}
\Delta q_8 = 3F-D
\end{equation}
where $F$ and $D$ are the antisymmetric and symmetric $SU(3)$ coupling 
constants of hyperon beta decays \cite{Bour,FD}. $\Delta q$ denotes the
first moment, i.e. the total polarization of each quark (or combination of
quarks), which is defined as:
\begin{equation}
\Delta q=\int^1_0 dx \Delta q(x).
\end{equation}
The $SU(3)_{flavour}$ non-singlet combinations are defined by:
\begin{eqnarray} \nonumber
\Delta q_8&=&(\Delta u+\Delta \bar{u})+
(\Delta d+\Delta \bar{d})-2(\Delta s+\Delta \bar{s}), \\
\Delta q_3&=&(\Delta u+\Delta \bar{u})-
(\Delta d+\Delta \bar{d}).
\end{eqnarray}
Assuming that the sea contribution for quarks and anti-quarks are equal,
first moments of above non-singlet combinations become:
\begin{eqnarray} \nonumber
\Delta q_8&=&\Delta u_v + \Delta d_v + 2\Delta S - 4 \Delta \bar{s} , \\
\Delta q_3&=&\Delta u_v - \Delta d_v - 2 \Delta\delta .
\end{eqnarray}
As we do not fix the first moment $\Delta q_3$ we are able to test the 
Bjorken sum rule \cite{BjSum}
\begin{equation}
\Delta q_3=F+D .
\end{equation}
We do not put $\Delta\delta(x,Q^2)=0$ (we distinguish $\Delta\bar{u}$
and $\Delta\bar{d}$), thus we are able to test $SU(2)_{isospin}$
breaking effects. Other first moments that can be calculated using the 
obtained integrated quark polarizations are:
\begin{eqnarray} \nonumber
&&\Delta\Sigma=\Delta u_v+\Delta d_v+2\Delta\bar{s}+2\Delta S\; ,\\
&&\Gamma^p_1=\frac{2}{9}\Delta u_v+\frac{1}{18}\Delta d_v+
\frac{1}{9}\Delta\bar{s}+\frac{5}{18}\Delta S-\frac{1}{6}\Delta\delta\; ,\\
\nonumber
&&\Gamma^n_1=\frac{1}{18}\Delta u_v+\frac{2}{9}\Delta d_v+
\frac{1}{9}\Delta\bar{s}+\frac{5}{18}\Delta S+\frac{1}{6}\Delta\delta\; ,
\end{eqnarray}

  The remaining 16 coefficients of eqs.~(\ref{eq10}) are determined by fitting 
the available data on the inclusive spin asymmetries for proton, neutron and 
deuteron targets and on the semi-inclusive spin asymmetries for the proton and 
deuteron target. The fit is performed assuming that the spin asymmetries do 
not depend on $Q^2$. Although the latter assumption is  not consistent with
theoretical predictions ($Q^2$-evolution of the numerator of 
eqs.~(\ref{eq0},\ref{eq4}) 
differs from $Q^2$-evolution of the denominator due to different polarized 
and unpolarized splitting  functions), it is  consistent with experimental 
observation [11--19].

  The results for the parameters in eq.~(\ref{eq10}) derived from the fit 
to data on inclusive and semi-inclusive spin asymmetries are presented below:
\begin{equation}
\begin{array}{llll}
A_u=0.175\;,& B_u=0.301\;,& C_u=2.010\;,& D_u=6.752\;, \\
A_d=-0.381\;,& B_d=0.083\;,& C_d=0.046\;,& D_d=-2.944\;, \\
A_s=-0.00052\;,& B_s=-0.007\;,& &  \\
A_S=0.026\;,& B_S=0.574\;,& C_S=-1.863\;,&  \\
A_{\delta}=0.002\;,& B_{\delta}=-0.289\;,& & \\
\end{array}
\end{equation}

\vskip 0.5cm
\begin{figure}[bht] 
\vskip 10cm\relax\noindent\hskip -0.5cm
       \relax{\includegraphics{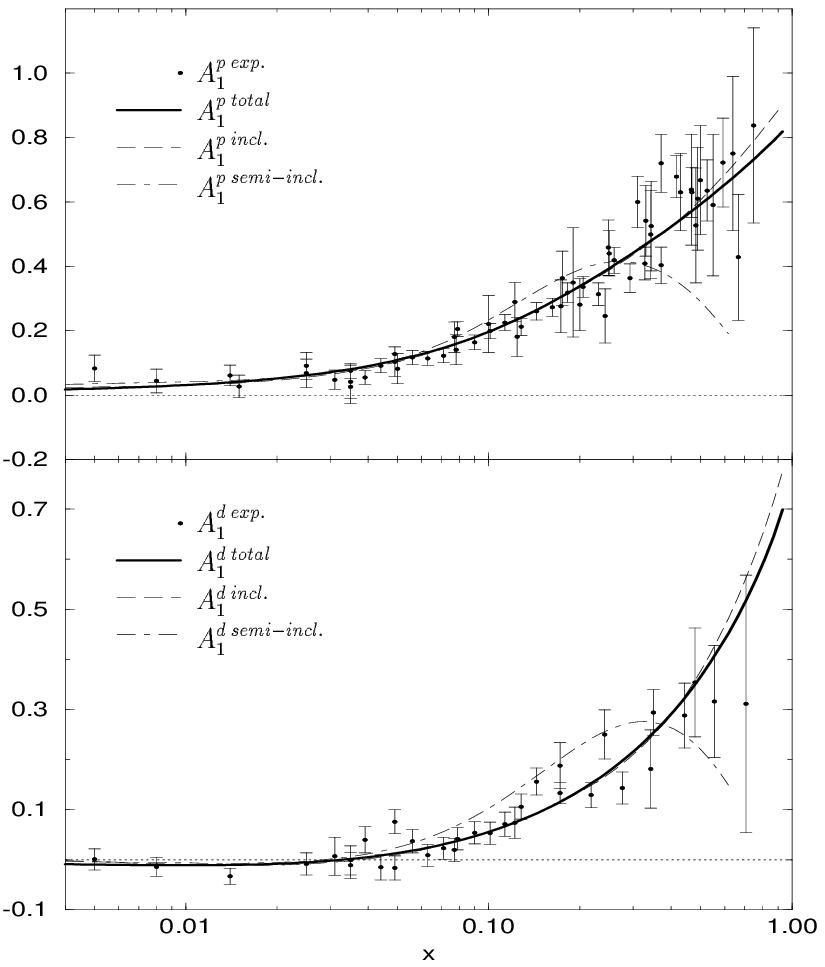}}
	\relax\noindent\hskip 8.5cm
       \relax{\includegraphics{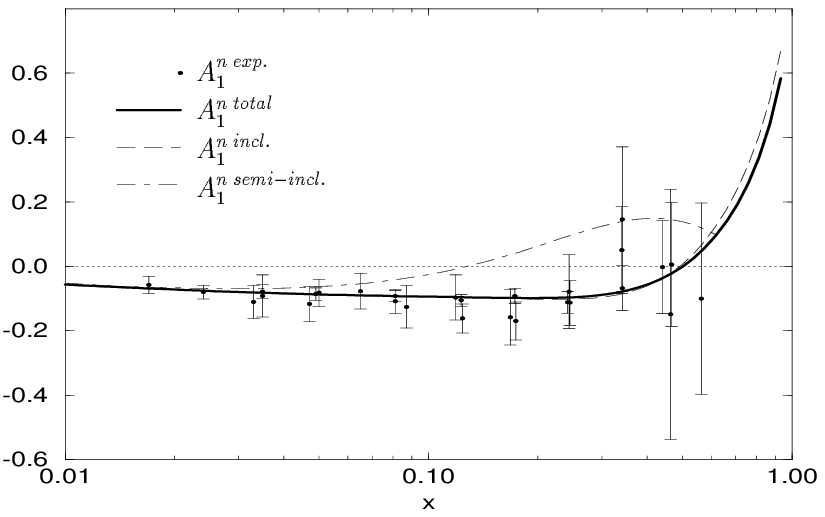}} 
\vskip -5cm\relax\noindent\hskip 8.5cm\begin{minipage}[t]{7.2cm}
\caption{\em The inclusive spin asymmetries $A_1^{N\;total}$ obtained from 
the total fit to the inclusive and semi--inclusive data, compared to all 
existing inclusive data points. $A_1^{N\;incl.}$ and $A_1^{N\;semi-incl.}$ are 
predictions which come out from the fit to data on inclusive and semi-inclusive
spin asymmetries, correspondingly.}
\label{fig1}
\end{minipage}
\vskip 3cm
\end{figure}

\begin{figure}[ht] 
\vskip 11cm\relax\noindent\hskip -0.5cm
       \relax{\includegraphics{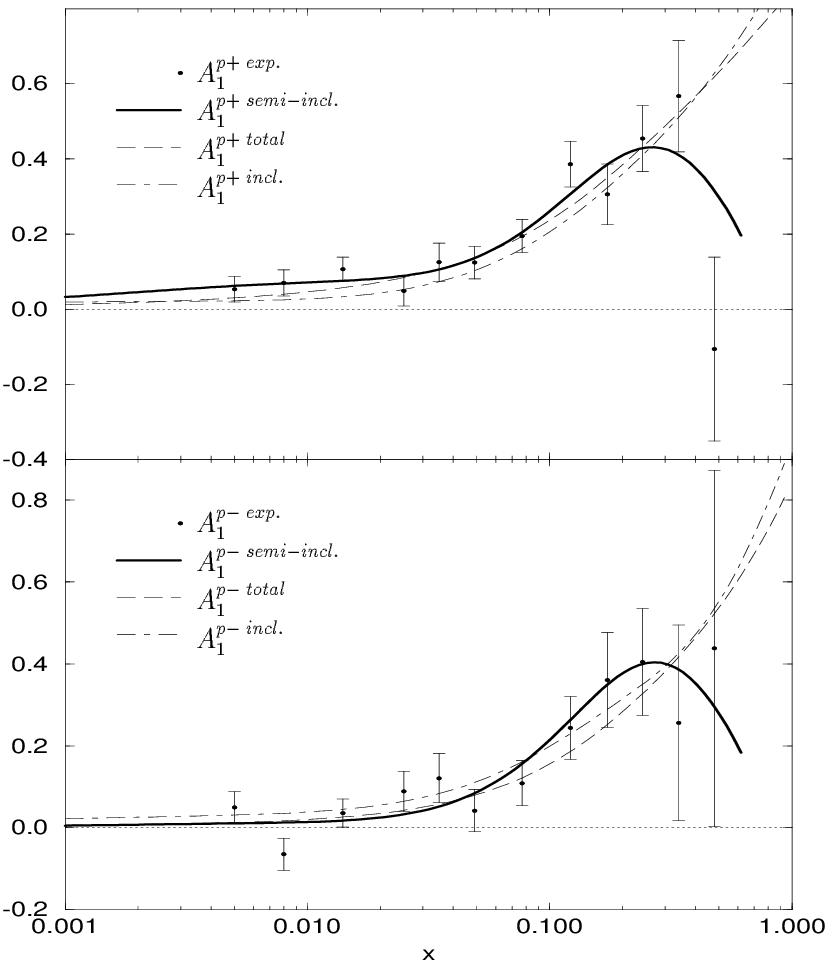}}
	\relax\noindent\hskip 8.5cm
       \relax{\includegraphics{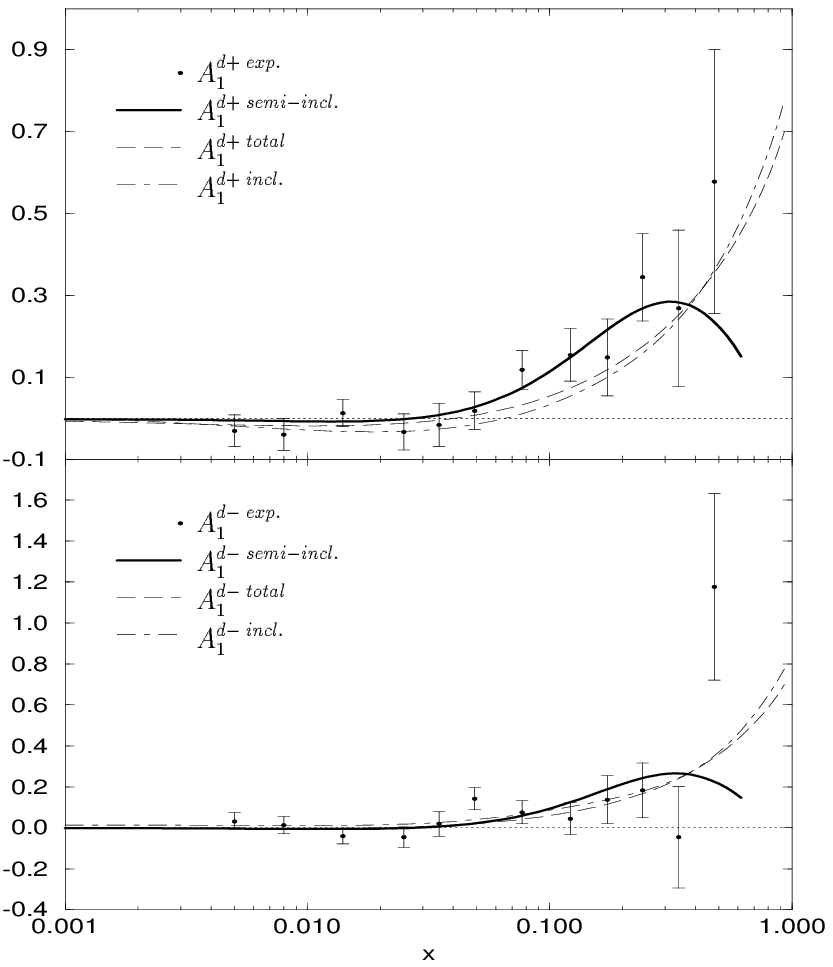}}
\caption{\em The semi-inclusive spin asymmetries obtained from the  
fit to the data on semi-inclusive spin asymmetries, compared to recent 
results presented by SMC \protect\cite{SMCsi97}. $A_1^{N\pm\;total}$ denotes 
semi-inclusive asymmetries obtained from the total fit to inclusive and 
semi-inclusive data. Predictions for semi-inclusive asymmetries calculated 
using distributions which come out from the fit to inclusive data are also 
presented ($A_1^{N\pm\;incl.}$). Note the last data point for $A_1^{p+}$, which
gives the largest contribution to $\chi ^2$, even for the semi-inclusive fit.}
\label{fig2}
\vskip 1cm
\end{figure}

\begin{figure}[bht] 
\vskip 5cm\relax\noindent\hskip -0.5cm
       \relax{\includegraphics{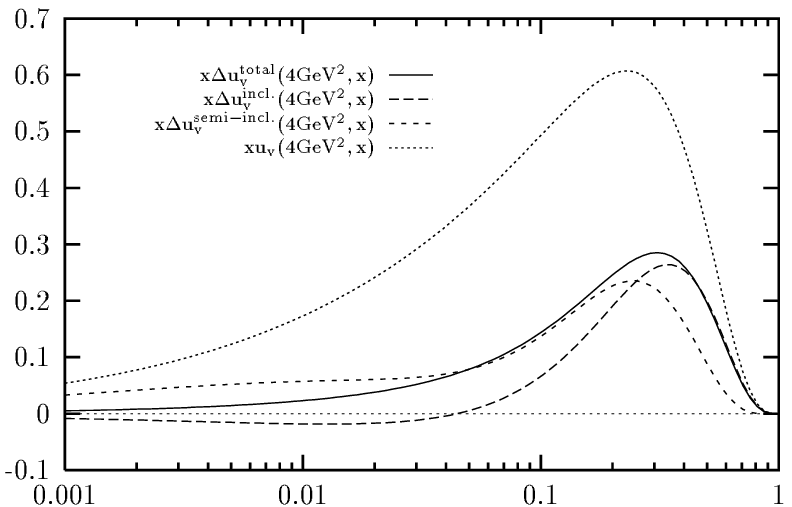}}
	\relax\noindent\hskip 8.5cm
       \relax{\includegraphics{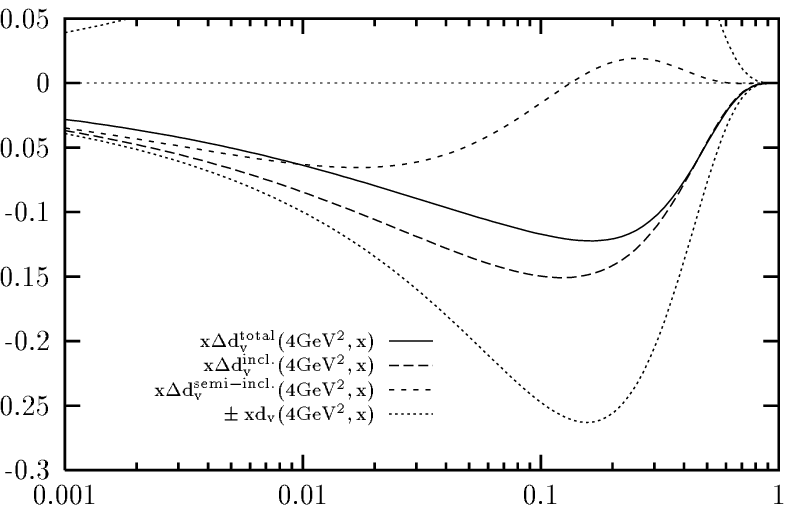}}
\vskip 5cm\relax\noindent\hskip -0.5cm
       \relax{\includegraphics{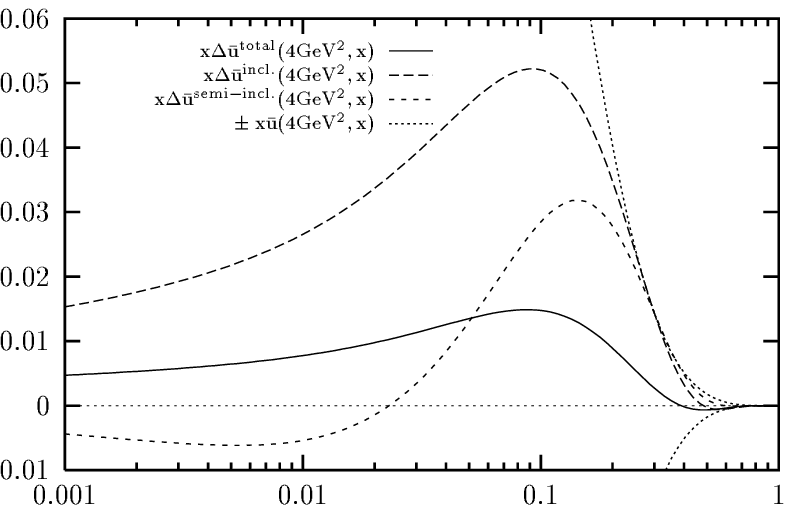}}
	\relax\noindent\hskip 8.5cm
       \relax{\includegraphics{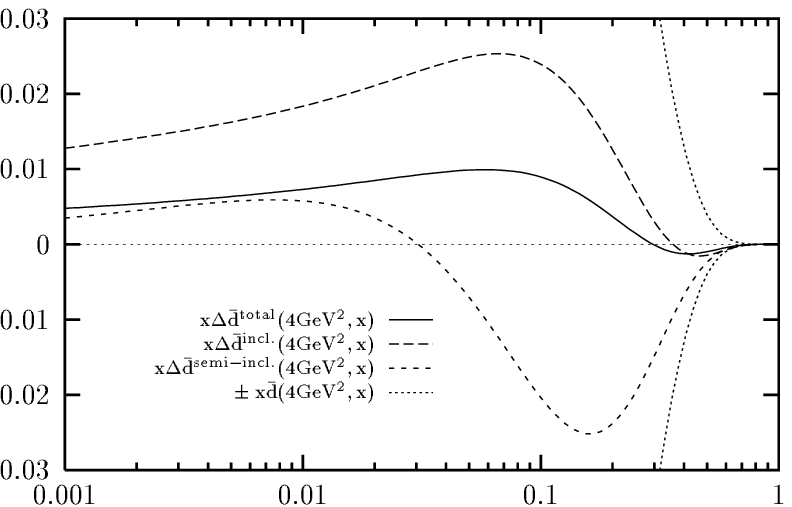}}
\vskip 5cm\relax\noindent\hskip -0.5cm
       \relax{\includegraphics{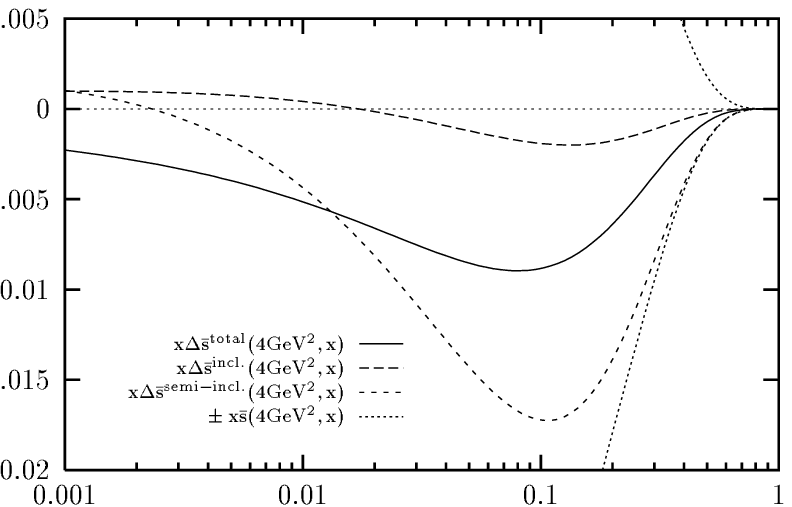}}
\vskip -5cm\relax\noindent\hskip 8.5cm\begin{minipage}[t]{7.2cm}
\caption{\em The distributions derived from the total fit and 
from fits to the data on inclusive and semi-inclusive spin 
asymmetries separately.}
\label{fig3}
\end{minipage}
\vskip 3cm
\end{figure}

For the fit we get $\chi^2=147$ for 157 d.o.f. (we use 123 data points from 
totally inclusive experiments and 48 data points from semi-inclusive
experiment), hence $\chi^2/N_{d.o.f.}=0.94$. Results for the inclusive
asymmetries for the proton, neutron and deuteron target are compared to the 
experimental data in Fig.~\ref{fig1}. The comparison of the {\em fitted} 
semi-inclusive spin asymmetries for production of positively and
negatively charged hadrons from the proton and deuteron target to experimental
points is given in Fig.~\ref{fig2}. Polarized quark distributions are
presented in Fig.~\ref{fig3}.

The values of the first moments of parton distributions are as follows:
\begin{equation}
\begin{array}{ll}
\Delta u_v=0.60\pm 0.01\;,& \Delta \bar{u}=0.08\pm 0.02\;, \\
\Delta d_v=-0.56\pm 0.01\;,& \Delta \bar{d}=0.07\pm 0.02\;, \\
\Delta \bar{s}=-0.042\pm 0.004\;.&  \\
\end{array}
\end{equation}
From these numbers one can evaluate
the values of the first moments of the structure functions and other 
combinations of the polarized quark parton distributions:
\begin{equation}
\begin{array}{ll}
\Delta u=0.68\pm 0.02\;,& \Delta d=-0.49\pm 0.02\;, \\ 
\Gamma^p_1=0.142\pm 0.002\;,& \Gamma^n_1=-0.057\pm 0.005\;, \\
\Delta\Sigma =0.26\pm 0.01\;,& \Delta q_3=1.19\pm 0.07\;, \\
\multicolumn{2}{l}{\Delta q_{sea}=
2(\Delta\bar{u}+\Delta\bar{d}+\Delta\bar{s})=0.22\pm 0.04\;.}
\end{array}
\end{equation}
In our model
quark contribution to the spin of the proton is dominated by the sea 
polarization. The contribution of each valence quark is almost the same
but has the opposite sign, hence valence quarks carry little of 
the spin of the proton. However all distributions are fitted to the data
points from the measured region, i.e. for $x>0.003$, and contributions beyond
the measured region ($\int^{0.003}_0 dx\, f(x)$) are questionable. Moreover
the low $x$ behaviour of the polarized quark parton distributions is 
determined by the unpolarized ones, therefore it is not consistent with the
Regge theory prediction. Though application of Regge theory is incompatible
with the pQCD \cite{lowx_nlo,lowx_bfr} it is interesting to compare
predictions of the Regge type behaviour for low $x$ to the predictions of
our model, especially as some of the considered quantities change rapidly
for $x<0.003$. For example $\int^{0.003}_0 dx\, \Delta d_v(x)=-0.113$
while $\int^1_{0.003} dx\, \Delta d_v(x)=-0.451$. It is the consequence of
the  power-like behaviour of the type $x^{-0.624}$. More dramatical
change can be observed for the non-strange sea quark distribution. The
contribution of the low $x$ region is $\int^{0.003}_0 dx\, \Delta
S(x)=0.072$ while for $x>0.003$ we get $\int^1_{0.003} dx\, \Delta
S(x)=0.079$. This is due to the fact, that $\Delta S(x)$ is more singular
for x tending to $0$ than any valence quark distribution. Multiplication of
the term $x^{-0.842}$ by the term $(\ln \frac{1}{x})^{0.037}$ enlarges the
value of integral over the range from $x=0$ to $x=0.003$, but not
significantly. The strange quark distribution behaviour for low x is the
most complex one. The term $x^{-0.5}$ is suppressed with the $(\ln
\frac{1}{x})^{-1.15}$ term but the multiplying term 
$e^{2.592 \sqrt{\ln\frac{1}{x}}}$ increases the contribution to the 
integral of the low $x$ region very fast. Finally an integration over $x$
below $0.003$ gives $-0.012$ to the total polarization of the quark $\bar{s}$
which is $-0.043$. 

Now we can compare our results to the more stable Regge
theory prediction.
The quantities integrated over region from $x=0.003$ to $x=1$ (integration
over the region covered by the experimental data plus extrapolation for higher
$x$) extrapolated to $x=0$, postulating the Regge type behaviour for all
quark parton distributions of a type $x^{-0.5}$, give:
\begin{equation}
\begin{array}{lllll}
\Delta u_v=0.61\;,& \Delta d_v=-0.54\;,&
\Delta \bar{u}=0.06\;,& \Delta \bar{d}=0.05\;,&
\Delta \bar{s}=-0.04\;,  \\
\Gamma^p_1=0.132\;,& \Gamma^n_1=-0.063\;, &
\Delta\Sigma =0.20\;,& \Delta q_3=1.17\;, &
\Delta q_{sea}=0.13\;.
\end{array}
\end{equation}

Our results on first moments of the proton and neutron structure functions
are in agreement with experimental results given in
Ref.\cite{SMCp}. Other  estimations in 
Ref.\cite{E142,E143p,E154,HERMES} are slightly smaller but our 
results are still consistent within two standard deviations. 
For the first moment of 
the deuteron structure function $g_1^d=\frac{1}{2}(g_1^p+g_1^n)(1-1.5p_D)$ we 
get  $\Gamma_1^d=0.039\pm 0.004$ (or $\Gamma_1^d=0.032$ if the Regge behaviour
for low x is assumed) what is in excellent agreement with results in 
Ref.\cite{E143d,SMCd}. For the purely non-singlet combination of the structure 
functions $(g_1^p-g_1^n)$, which in our model is 
$\Gamma_1^{p-n}=\frac{1}{6}\Delta q_3$, we obtain $\Gamma_1^{p-n}=0.198$
($0.195$ assuming Regge behaviour). This value is in good 
agreement with the $O(\alpha_s^3)$ \cite{GorLar} prediction 0.188 
($\alpha_s(M_Z^2)=0.109$). The non-singlet combination is expected to be 
less sensitive to the low $x$ shape than its singlet counterpart
\cite{2logs}. Similarly, we observe that the value of $\Delta q_3$ varies
between 1.19 and 1.17 while $\Delta u+\Delta d=\Delta u_v+\Delta
d_v+2\Delta S$ changes from 0.34 in our model to 0.28 for Regge-type
behaviour. We obtain a quite large and positive
non-strange sea polarization and the whole sea polarization alike
($2\Delta\bar{s}=-0.08$ seems to be reasonable). Finally, $\Delta\Sigma
=0.26\,(0.20)$ is consistent with existing determinations.

\begin{figure}[ht] 
\vskip 5cm\relax\noindent\hskip -0.5cm
       \relax{\includegraphics{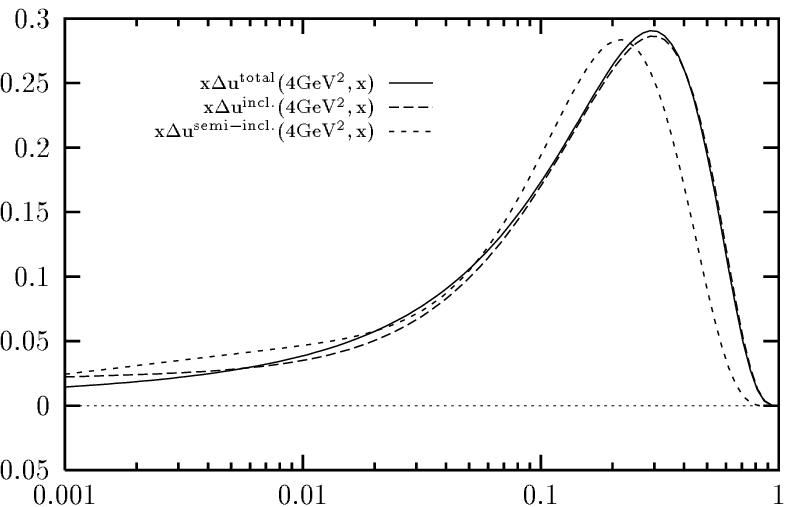}}
	\relax\noindent\hskip 8.5cm
       \relax{\includegraphics{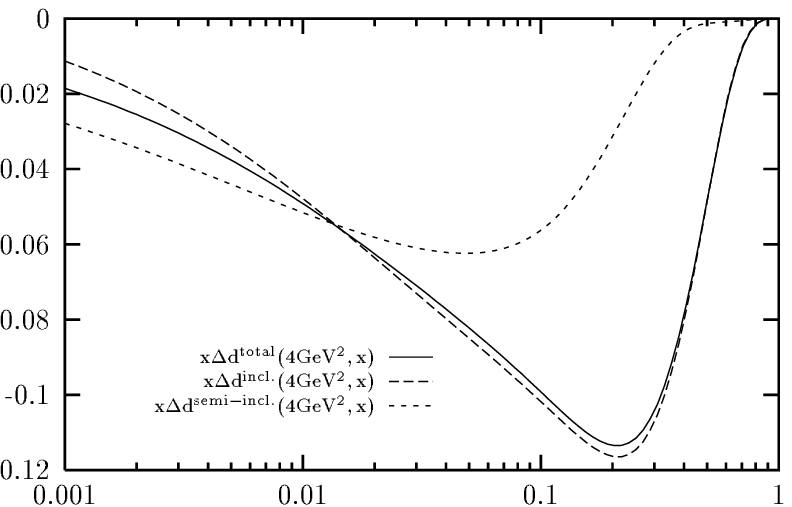}}
\caption{\em The whole $\Delta d$ and $\Delta u$ distributions derived from 
the total fit and from fits to the data on inclusive and semi-inclusive 
spin asymmetries separately.}
\label{fig4}
\end{figure}

Performing fits to the inclusive and semi-inclusive data separately we can
test the impact of each type of data on the total fit. When we use our model
to make a fit to the data on inclusive spin asymmetries solely, 
we get $\chi^2=95$,
nearly equal to 96.6, which is the contribution of the inclusive data points to 
the $\chi^2=147$ of the total fit ($\chi^2/N_{d.o.f.}=0.87$ is better then
in the total case $\chi^2/N_{d.o.f.}=0.94$).
It can be observed in Fig.\ref{fig1}, that inclusive spin asymmetries derived
from both types of fits do not differ much from each other in our model. 
Moreover results
on the integrals over $0.003<x<1$ of  the structure functions and singlet
or non-singlet combinations of quark distributions are very close to those
obtained in the total fit. In details: $\Delta\Sigma =0.27$, 
$\Gamma_1^p=0.126$, $\Gamma_1^n=-0.050$, $\Delta q_3 =1.05$. Although the total
polarization of quarks of a certain flavour (i.e. 
$\Delta u(d,s)=\Delta u(d)_v + 2\Delta\bar{u}(\bar{d},\bar{s})$) does not 
vary much, the division between valence and sea quarks differs from the 
total fit considerably. For example, in $0.003<x<1$ region, we get 
$\Delta u_v=0.336$, $\Delta d_v=-0.565$. 
The reason is, that the valence and sea quark distributions of the same 
flavour have the same weight (electric charge squared) in the inclusive
spin asymmetry (eq. (\ref{eq3})). Hence the asymmetry is sensitive only to the
whole $\Delta q$ distributions but not to valence and sea quark distributions
separately. The distributions $\Delta u$ and $\Delta d$ derived from the total
fit and the fit to inclusive data have the same shape, which can be seen in 
Fig. \ref{fig4}, whereas splits between valence and sea quark distributions
are different in both cases (compare Fig. \ref{fig1}).

The comparison to the similar analysis (performed in Ref.\cite{BT}), which uses 
the MRS parametrization as an input shows us the influence of choice of input
parametrization on distributions and first moments. There is almost no 
difference between first moments of distributions for quarks of a certain
flavour obtained in Ref.\cite{BT} and in this analysis. Corresponding 
results are: $\Delta u=0.76$ (0.70 assuming Regge behaviour), 
$\Delta d=-0.52$ (-0.37), $\Delta{s}=-0.07$(-0.07). There is a significant 
difference in division between valence and sea quarks. What is most important,
the total polarization of the sea quarks changes its sign. In Ref.\cite{BT}
$\Delta q_{sea}=-0.18$(-0.22) whereas we have obtained 
$\Delta q_{sea}=0.22$(0.13).

Parameterizations obtained using inclusive and semi-inclusive data give a good
description of the semi-inclusive asymmetries, as can be seen in 
Fig. \ref{fig4} (the contribution of 48 semi-inclusive data points to 
$\chi^2=147$ of the total fit amounts 50.4, where the data point for 
$A_1^{p+}$ at $x=0.48$ gives the biggest part). Still, fit performed 
using only semi-inclusive data leads to different set of coefficients of 
distribution functions and other integrated results, as is seen in 
Figs. \ref{fig1}, \ref{fig4}. This is mainly caused by the data point for 
$A_1^{p+}$ at $x=0.48$, which makes that semi-inclusive  asymmetries 
go below the total and inclusive predictions (Fig. \ref{fig2}). Absence of 
this point would improve the result of comparison to the experiment and
change high $x$ behaviour of the fitted semi-inclusive asymmetries.

We get for the fit with semi-inclusive data points only $\chi^2=39.3$ 
($\chi^2/N_{d.o.f.}=1.2$). The integrated quantities are:
\begin{equation}
\begin{array}{lll}
\Delta u_v=0.68\pm 0.01\;,& \Delta \bar{u}=0.02\pm 0.05\;, &
\Delta u=0.70\pm 0.05\;,\\
\Delta d_v=-0.32\pm 0.05\;,& \Delta \bar{d}=-0.02\pm 0.07\;, &
\Delta d=-0.34\pm 0.09\;,\\
\Delta \bar{s}=-0.035\pm 0.004\;,&& \\
\Gamma^p_1=0.137\pm 0.003\;,& \Gamma^n_1=-0.042\pm 0.018\;, &
\Gamma^d_1=0.044\pm 0.010\;, \\
\Delta\Sigma =0.30\pm 0.04\;,& \Delta q_3=1.08\pm 0.21\;, &
\Delta q_{sea}=-0.06\pm 0.17\;.
\end{array}
\end{equation}
There is no important difference between above first moments and integrals
obtained assuming Regge  behaviour for low $x$. These results are in a good 
agreement with experimental estimations \cite{SMCsi96,SMCsi97}. The
presence of various weights in the semi-inclusive spin asymmetries (eq.
(\ref{eq3})) induces that division of $\Delta u$ and $\Delta d$ between
valence and sea parts is no longer strongly model dependent. 
Also differences among sea
quarks of different flavours are emphasized. 
In Fig. \ref{fig1}, one sees
that the parametrizations obtained using only semi-inclusive data points
give inclusive asymmetries too far from experimental data points. If
we compute $\chi^2$ for all data points of both types with the obtained
distributions we get 505 what is an unacceptable value. The substantial
part comes from the E154 data for the neutron target, mainly due to the
differences in the whole $\Delta d$ distribution obtained from
semi-inclusive and total fits, as can be seen in Fig. \ref{fig4}.

We have performed an analysis of the world data on polarized deep inelastic
scattering, inclusive and semi-inclusive, assuming that $\Delta\bar{u}\not
=\Delta\bar{d}$, i.e. $\Delta\delta\not=0$. But in the inclusive case only 
whole quark distributions of a certain flavour are distinguished,
 i.e. have different weights in the asymmetry. As $\Delta u(d)=
\Delta u(d)_v+\Delta S \mp \Delta\delta$, putting $\Delta\delta\not=0$
gives two additional coefficients, very weakly constrained. We have
performed also a fit to the inclusive data only, putting $\Delta\delta=0$
and we have got almost the same value of $\chi^2$ as before. Although it
is possible to obtain the information about difference between
$\Delta\bar{u}$ and $\Delta\bar{d}$ without taking into account
semi-inclusive data, the results are very poorly constrained by the data.

The situation is better in the semi-inclusive case where all of the 
distributions (valence and sea separately) appear in the semi-inclusive
spin asymmetry with different weights. Hence, up till now, there are
24 data points\footnote{for the deuteron target $\Delta\delta$ does not 
appear in  the formula for the semi-inclusive spin asymmetry.} constraining
the coefficients of the $\Delta\delta$ distribution. Performing a fit 
to the semi-inclusive data with $\Delta\delta=0$ we have got a slightly 
worse value of $\chi^2$ then in the case with $\Delta\delta\not=0$.

The inclusion of available semi-inclusive data to the analysis of the 
inclusive events gives more stable results. Using data on 
semi-inclusive spin asymmetries we can distinguish  valence and sea quarks 
distributions of the same flavour as well as $\Delta\bar{u}$ and 
$\Delta\bar{d}$. However parametrization obtained as the best fit to the 
semi-inclusive data gives the unacceptable description of the inclusive 
ones, mainly due to the differences in the  $\Delta d$ distribution.
Hence, we have got no perfect consistence of inclusive and semi-inclusive 
results in our model. 
Additional data from semi-inclusive experiments using $^3 He$ target can 
reverse the situation. The next step in the analysis, i.e. addition of 
the $Q^2$-dependence of distribution functions can also improve an agreement 
of our model with an experiment.

Our analysis shows that the result which gives the polarization of the sea 
quarks depends strongly on used parametrization of the polarized parton
distributions.

\vfill



\pagebreak

\bibliographystyle{unsrt}

\def\jou#1#2#3#4{{#1} {\bf #2}, #3 (#4)}
\def\etal{et al.}

\def\NCA{{\em Nuovo Cimento}}
\def\NIM{{\em Nucl. Instrum. Methods}}
\def\NIMA{{\em Nucl. Instrum. Methods} {\bf A}}
\def\NPB{{\em Nucl. Phys.} {\bf B}}
\def\PLB{{\em Phys. Lett.}  {\bf B}}
\def\PRL{{\em Phys. Rev. Lett.}}
\def\PR{{\em Phys. Rev.}}
\def\PRD{{\em Phys. Rev.} {\bf D}}
\def\PRC{{\em Phys. Rev.} {\bf C}}
\def\ZPC{{\em Z. Phys.} {\bf C}}
\def\APPB{{\em Acta Physica Polonica} {\bf B}}
\def\PRep{{\em Phys. Rep.}}

\end{document}